\documentclass[smallabstract,smallcaptions]{dccpaper}

\usepackage{epsfig}
\usepackage{citesort}
\usepackage{amsmath}
\usepackage{amssymb}
\usepackage{color}
\usepackage{url}

\usepackage{graphicx}

\newlength{\figurewidth}
\newlength{\smallfigurewidth}

\begin{document}

\title{\large
\textbf{Information profiles for DNA pattern discovery}}

\author{%
Armando J. Pinho, Diogo Pratas, and Paulo J. S. G. Ferreira\\[0.5em]
{\small\begin{minipage}{\linewidth}\begin{center}
\begin{tabular}{ccc}
\multicolumn{3}{c}{IEETA / Dept of Electronics, Telecommunications and
Informatics}\\
\multicolumn{3}{c}{University of Aveiro, 3810--193 Aveiro, Portugal}\\
\multicolumn{3}{c}{\url{ap@ua.pt} --- \url{pratas@ua.pt} --- \url{pjf@ua.pt}}
\end{tabular}
\end{center}\end{minipage}}
}

\maketitle
\thispagestyle{empty}

\begin{abstract}
Finite-context modeling is a powerful tool for compressing and hence for
representing DNA sequences. We describe an algorithm to detect genomic
regularities, within a blind discovery strategy. The algorithm uses
information profiles built using suitable combinations of finite-context
models. We used the genome of the fission yeast
\textit{Schizosaccharomyces pombe} strain 972 h$^-$ for illustration,
unveiling locations of low information content, which are usually
associated with DNA regions of potential biological interest.
\end{abstract}

%
\Section{Introduction}
\label{sec:intro}
Graphical representations of DNA sequences are a handy way of quickly
finding regions of potential interest. This has been a topic addressed
using various approaches (see, for example, \cite{Schneider-1990a,%
Jeffrey-1990a,Oliver-1993a,Goldman-1993a,Jensen-1999a,Deschavanne-1999a,%
Crochemore-1999a,Troyanskaya-2002a,Fertil-2005a,Vinga-2007a}), some of
them relying on information theoretical principles. Both global and
local estimates of the randomness of a sequence provide useful
information, but both also have shortcomings. Global estimates
do not show how the characteristics change along the sequence.
Local estimates fail to take into consideration the global properties
of the sequence. This latter drawback was addressed by Clift \textit{et al.}
\cite{Clift-1986a} using the concept of sequence landscape, plots
displaying the number of times oligonucleotides from the target
sequence occur in a given source sequence. If the target and source
sequences coincide, then the landscape provides information about
self-similarities (repeats) of the target sequence.

The sequence landscapes of Clift \textit{et al.} \cite{Clift-1986a}
have been a first attempt to display local information, taking into
account global characteristics of the sequence. This idea was pursed
by Allison \textit{et al.} \cite{Allison-2000a} using XM, a model
that considers a sequence as a mixture of regions with little structure
and regions that are approximate repeats. With this statistical model,
they have produced information sequences, which quantify the amount
of surprise of having a given base at a given position, knowing the
remaining of the sequence. When plotted, one of these information
sequences provides a quick overview of certain properties of the
original symbolic sequence, allowing for example to easily identify
zones of rich repetitive content \cite{Stern-2001a,Cao-2007a,Dix-2007a}.

The information sequences of Allison \textit{et al.} \cite{Allison-2000a}
are deeply related to data compression. The role of data
compression for pattern discovery in DNA sequences was initially
pointed out by Grumbach \textit{et al.} \cite{Grumbach-1993a} and,
since then, it has been pursued by other researchers (e.g. \cite{Rivals-1997a,%
Stern-2001a}). In fact, the algorithmic information content of a sequence
is the size, in bits, of the shortest description of the sequence.

In this paper, we propose using combinations of several finite-context
models, each of a different depth, for building \textit{information
profiles}. Such models have been shown to adequately capture the statistical
properties of DNA sequences \cite{Pinho-2009b,Pratas-2011a,Pinho-2011a,%
Pinho-2011e} but are direction-dependent, i.e., the results depend on
which direction the DNA sequence is processed. We remove this directional
dependency by combining the amount of information that a certain DNA
base carries in each processing direction.

The information profiles are found using an algorithm based on
finite-context models that needs time proportional to the length of the
sequence. We present a proof-of-concept study of
the potential of information profiles in genome analysis, namely,
for detecting genomic structural and functional regularities. We
uncover genomic regularities on a large-scale, such as, centromeric
and telomeric regions of a chromosome, or transposable elements. In
this context, we use the genome of the fission yeast
\textit{Schizosaccharomyces pombe} strain 972~h$^-$ as case-study.

%
\Section{Building the information profiles}
Finite-context models are probabilistic models based on the
assumption that the information source is Markovian, i.e., that the
probability of the next outcome depends only on some finite number of
(recent) past outcomes referred to as the context.
The proposed approach is based on a mixture of finite-context
models. We assign probability estimates to each symbol in
$\mathcal{A} = \{\mathrm{A},\mathrm{C},\mathrm{G},\mathrm{T}\}$,
regarding the next outcome,
according to a conditioning context computed over a finite and fixed
number $k > 0$ of past outcomes $x_{n-k+1..n} = x_{n-k+1}\dots x_{n}$
(order-$k$ finite-context model with $|\mathcal{A}|^k$ states).

The probability estimates $P(x_{n+1} | x_{n-k+1..n})$ are calculated using
symbol counts that are accumulated while the sequence is processed,
making them dependent not only on the past $k$ symbols, but also on
$n$. We use the estimator
\begin{equation} \label{eq:pe}
P(s | x_{n-k+1..n}) = \frac{C(s | x_{n-k+1..n}) + \alpha}
{C(x_{n-k+1..n}) + |\mathcal{A}|\alpha},
\end{equation}
where $C(s | x_{n-k+1..n})$ represents the number of times that, in the
past, symbol $s$ was found having $x_{n-k+1..n}$ as the conditioning
context and where
\begin{equation}
C(x_{n-k+1..n}) = \sum_{a \in {\mathcal A}} C(a | x_{n-k+1..n})
\end{equation}
is the total number of events that has occurred so far in association
with context $x_{n-k+1..n}$. Parameter $\alpha$ allows balancing between
the maximum likelihood estimator and a uniform distribution (when the
total number of events, $n$, is large, it behaves as a maximum likelihood
estimator). For $\alpha = 1$,~(\ref{eq:pe}) reduces to the well-known
Laplace estimator.

The per symbol information content average provided by the finite-context
model of order-$k$, after having processed $n$ symbols, is given by
\begin{equation}
H_{k,n} = -\frac{1}{n}\sum_{i=0}^{n-1} \log_2 P(x_{i+1} | x_{i-k+1..i})
\label{entropy}
\end{equation}
bits per symbol. When using several models
simultaneously, the $H_{k,n}$ can be viewed as measures of the performance
of those models until that instant. Therefore, the probability estimate
can be given by a weighted average of the probabilities provided by each
model, according to
\begin{equation}\label{eq:mixture}
P(x_{n+1}) = \sum_k P(x_{n+1} | x_{n-k+1..n})\;w_{k,n},
\end{equation}
where $w_{k,n}$ denotes the weight assigned to model $k$ and
\begin{equation}
\sum_k w_{k,n} = 1.
\end{equation}
Our modeling approach is based on a mixture of probability estimates.
In order to compute the probability estimate for a certain symbol, it is
necessary to combine the probability estimates given by~(\ref{eq:pe})
using~(\ref{eq:mixture}). The weight assigned to model $k$ can be
computed according to
\begin{equation}
w_{k,n} = P(k | x_{1..n}),
\label{eq:wkn}
\end{equation}
i.e., by considering the probability that model $k$ has generated
the sequence until that point. In that case, we would get
\begin{equation}
w_{k,n} = P(k | x_{1..n}) \propto P(x_{1..n} | k) P(k),
\end{equation}
where $P(x_{1..n} | k)$ denotes the likelihood of sequence $x_{1..n}$
being generated by model $k$ and $P(k)$ denotes the prior probability
of model $k$. Assuming
\begin{equation}
P(k) = \frac{1}{K},
\end{equation}
where $K$ denotes the number of models, we also obtain
\begin{equation}
w_{k,n} \propto P(x_{1..n} | k).
\end{equation}
Calculating the logarithm we get
\begin{subequations}
\begin{gather}
\log_2 P(x_{1..n} | k) = \log_2 \prod_{i=1}^n P(x_i | k, x_{1..i-1}) =\\
  = \sum_{i=1}^n \log_2 P(x_i | k, x_{1..i-1}) =
  \sum_{i=1}^{n-1} \log_2 P(x_i | k, x_{1..i-1}) + \log_2 P(x_n | k,x_{1..n-1}),
    \label{eq:lc2}
\end{gather}
\label{eq:lc}
\end{subequations}
which is related to the number of bits that would be required by model
$k$ for representing the sequence $x_{1..n}$. It is, therefore, the
accumulated measure of the performance of model $k$ until instant $n$.

DNA sequences are known to be non-stationary. Due to this, the performance
of a model may vary considerably from region to region of the sequence.
In order to extract the best possible performance from each model, we
adopted a progressive forgetting mechanism. The idea is to allow each model
to progressively forget the distant past and, consequently, to give more
importance to recent outcomes. Therefore, we write a modified version of
(\ref{eq:lc2}) as
\begin{equation}
\log_2 p_{k,n} = \gamma \log_2 p_{k,n-1} + \log_2 P(x_n | k, x_{1..n-1}),
\label{eq:forgetting}
\end{equation}
where $\gamma \in [0, 1)$ dictates the forgetting factor and
$\log_2 p_{k,n}$ represents the estimated number of bits that would
be required by model $k$ for representing the sequence $x_{1..n}$
(we set $p_{k,0} = 1$), taking into account the forgetting mechanism.

Removing the logarithms, we can rewrite (\ref{eq:forgetting}) as
\begin{equation}
p_{k,n} = p_{k,n-1}^{\gamma} P(x_n | k, x_{1..n-1})
\end{equation}
and, finally, set the weights to
\begin{equation}
w_{k,n} = \frac{p_{k,n}}{\sum_{k}p_{k,n}}.
\end{equation}

This probabilistic model yields an estimate of the probability
of each symbol in the DNA sequence, and as such it allows us to quantify the
degree of randomness or surprise along one direction of the sequence.

%
\Section{Results and Discussion}
For illustration, we used the \textit{S. pombe} genome (uid 127),
obtained from the National Center for Biotechnology Information (NCBI)%
\footnote{\url{ftp://ftp.ncbi.nlm.nih.gov/genomes/.}}. The profiles
are the result of the combination of eight finite-context models
with context depths of 2, 4, 6, 8, 10, 12, 14 and 16. Probabilities
were estimated with $\alpha=1/20$ in Eq.~\ref{eq:pe} for the larger
contexts of $k=$14 and $k=$16. For clarity, the full chromosome
profiles result from low-pass filtering with a Blackman window of
$1,001$ bases and sampling every 20 bases.

Chromosomes are processed
both in the downstream, or direct (5'$\rightarrow$3'), and upstream,
or reversed (3'$\rightarrow$5'), directions. This dual processing
aims at eliminating the directionality bias introduced when only one
of the two possible directions is taken into consideration.
Therefore, the information content of each DNA base is calculated
by running the statistical model in one direction, then in the other
direction, and finally by taking the smallest value obtained.

\begin{figure}[t]
\centerline{\includegraphics[width=15cm]{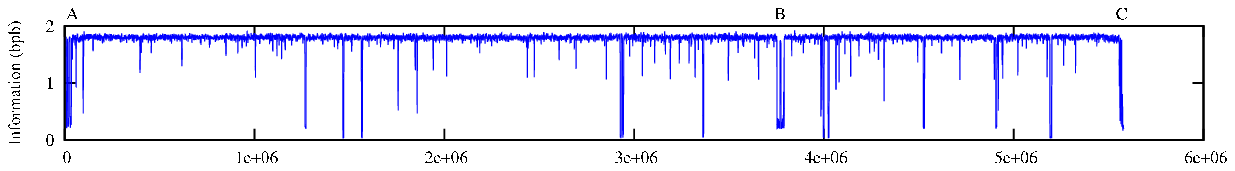}}
\vspace{1.0ex}
\centerline{\includegraphics[width=15cm]{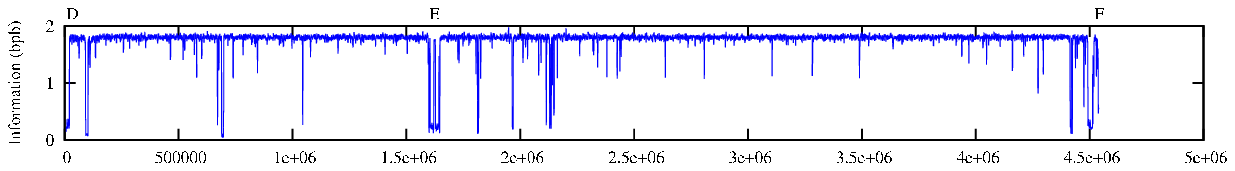}}
\vspace{1.0ex}
\centerline{\includegraphics[width=15cm]{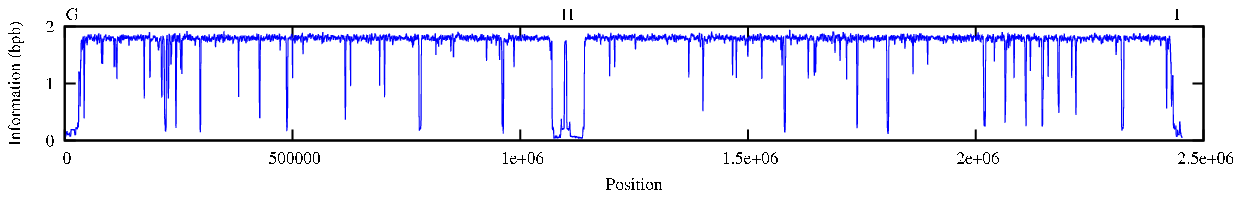}}
\vspace{0.5ex}
\caption{Plots of the information content for chromosome~I (first row),
chromosome~II (second row) and chromosome~III (third row) of \textit{S. pombe}.
The information profiles were obtained by processing the sequences in
both directions, and then choosing the minimum information value in each
direction. For better visualization, low-pass filtering with a Blackman
window of $1,001$ bases was applied to the profiles.}
\label{IndividualSpombe}
\end{figure}

\begin{figure}[t]
\centerline{\includegraphics[width=10cm]{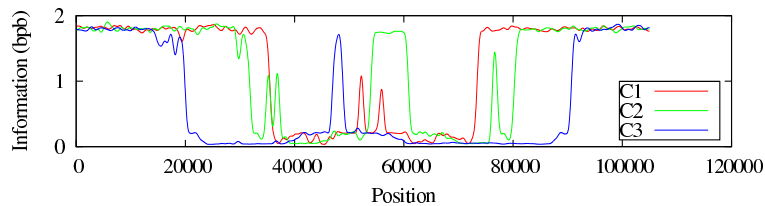}}
\vspace{0.5ex}
\caption{Plot of the information content of the centromeres of chromosome~I
(C1), II~(C2) and~III~(C3).}
\label{ALLCENT}
\end{figure}

We have computed the information profiles for each of the three
chromosomes (Fig.~\ref{IndividualSpombe}). There are locations of
low information content which are associated with DNA regions of
biological interest, such as telomeric and centromere regions.
We have marked with letters A, C, D, F, G and I the telomeric regions
and with letters B, E and H the centromere regions. They also allow
to identify the long arm (q) and short arm (p) on each chromosome.
In Fig.~\ref{ALLCENT}, we display a zoomed view of the centromeres,
revealing that their size varies inversely with the length of the
respective chromosome.

In general, low-information regions are associated with the presence
of repetitive sequences. For example, chromosome III has more and often
more prominent low-information regions than chromosomes I and II, which
is in compliance with some properties of this chromosome concerning
repetitive structures, such as, the presence of tandem rDNA repeats
\cite{Wood-2006a} or the density of transposable element remnants
in this chromosome being twice that of chromosomes I and II \cite{Wood-2002a}.

\begin{figure*}[t]
\centerline{\includegraphics[width=15.9cm]{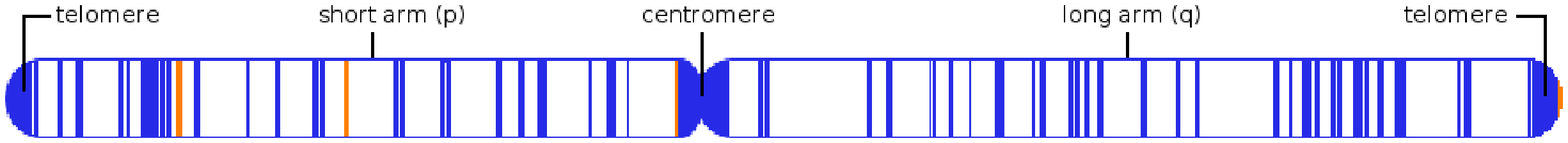}}
\vspace{1.0ex}
\centerline{\includegraphics[width=15cm]{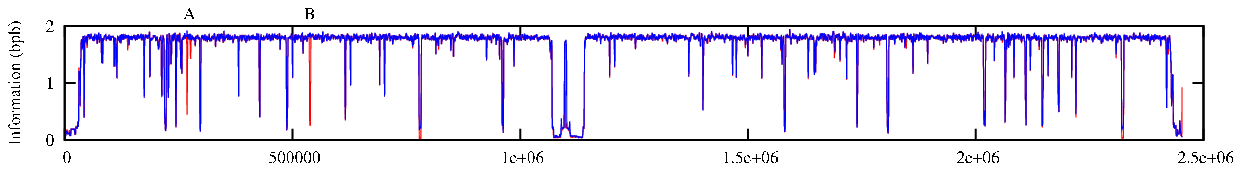}}
\vspace{1.0ex}
\centerline{\includegraphics[width=15cm]{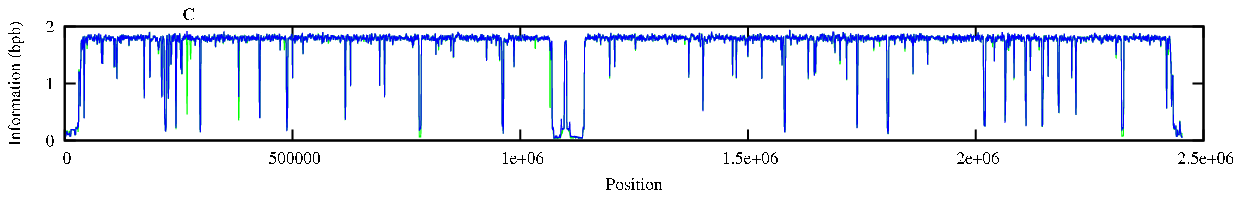}}
\vspace{0.5ex}
\caption{Information content of chromosome~III of \textit{S. pombe}.
The first row shows a representation for chromosome~III and their
long repetitive zones. The second row shows chromosome~III (blue)
with information added from chromosome~I (green). The third row shows
chromosome~III (blue) with information added from chromosome~II (red).}
\label{CatSpombe}
\end{figure*}

We have also performed an inter-chromosomal study. We concatenated
chromosome~I with chromosome~III and ran the algorithm from left to
right and from right to left, picking the lowest information content
values of both, base by base. A similar process has been done substituting
chromosome~I by II. Figure~\ref{CatSpombe} shows only the information
profile of chromosome~III calculated taking into account the
statistics of chromosome~I, second row, and chromosome~II, third row. 
We can see important regions marked with the letters A, B and C. The region 
marked with letter B contains the $2,529$ bases of gene eft202 (from base 
$537,326$ to $539,854$ in chromosome~III).
This gene is also present in chromosome~I, named eft201, located from
base $2,907,701$ to $2,910,229$, and has $\sim$99\% sequence
similarity to gene eft202.

\begin{figure}[t]
\centerline{\includegraphics[width=10cm]{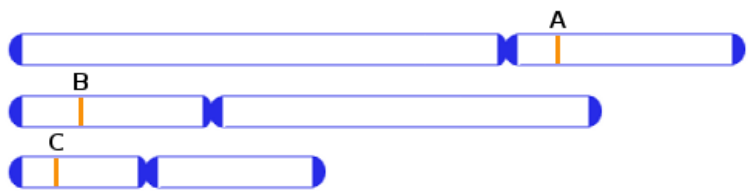}}
\vspace{0.5ex}
\caption{Illustration of the three chromosomes of \textit{S. pombe}
genome marked with genes ef1a-b (A), ef1a-c (B) and ef1a-a (C).}
\label{ilustrationChromos}
\end{figure}

The region marked with letter~A in Fig.~\ref{CatSpombe} indicates a
region in chromosome~III (gene ef1a-a) that is highly similar ($\sim$98\%
sequence similarity) with a region of chromosome~I (gene ef1a-b). Although
not included here, we found also an identical degree of similarity with
gene ef1a-c in chromosome~II. Fig.~\ref{ilustrationChromos} illustrates
the relative position of these genes, where letter~A marks a region from
base $4,095,202$ to $4,096,584$ ($1,383$ bases, chromosome~I), letter~B
refers from base $626,106$ to $627,488$ ($1,383$ bases, chromosome~II),
and letter~C from base $268,097$ to $269,479$ ($1,383$ bases, chromosome~III).

%
\Section{Conclusions}
We described an algorithm to detect genomic regularities within a
\textit{blind} discovery strategy. This algorithm uses information profiles
built using an efficient DNA sequence compression method. The results
described support our claim that information profiles provide
a valuable discovery tool for genome-wide studies. In fact,
the accurate matching of the low-information regions to annotated
repetitive genomic structures, such as the centromeric and telomeric
regions of a chromosome, proves information profiles
may be useful in \textit{de novo} discovery of large-scale genomic
regularities. Clearly, it is not possible to infer the genomic sequence
\textit{per se} from the information profiles, or the location of genomic
regularities within base pair resolution. However, it is possible to
discover the presence of regularities on a genome-wide scale, which may
be useful for an exploratory genome analysis or for genome comparisons.

Our algorithm relies on the efficient probabilistic modeling of the
genomic sequence based on finite-context models. The approach is
sufficiently flexible and powerful to enable addressing various
biological questions and quickly obtaining the corresponding information
profiles for a first-hand assessment. Indeed, the creation of
information profiles does not require high performance computational
facilities. Building an information profile requires a
computation time that depends only linearly on the size of the sequence.
For example, the information profile of a human chromosome can be
created in a laptop computer in just a few minutes.
Moreover, the amount of computer memory required does not depend on
the size of the sequence, but only on the depth of the finite
context models used for modeling the sequence.

%
\Section{Acknowledgements}
This work was supported in part by FEDER through the Operational Program
Competitiveness Factors - COMPETE and by National Funds through FCT -
Foundation for Science and Technology, in the context of the projects
FCOMP-01-0124-FEDER-022682 (FCT reference PEst-C/EEI/UI0127/2011) and
Incentivo/EEI/UI0127/2013.

\Section{References}
\bibliographystyle{IEEEtran}

\end{document}